%% file: main.tex
\documentclass[manuscript,screen]{acmart}

\AtBeginDocument{%
  \providecommand\BibTeX{{%
    \normalfont B\kern-0.5em{\scshape i\kern-0.25em b}\kern-0.8em\TeX}}}

\copyrightyear{2021}
\acmYear{2021}
\setcopyright{acmlicensed}\acmConference[CHI '21]{CHI Conference on Human Factors in Computing Systems}{May 8--13, 2021}{Yokohama, Japan} \acmBooktitle{CHI Conference on Human Factors in Computing Systems (CHI '21), May 8--13, 2021, Yokohama, Japan}
\acmPrice{15.00}
\acmDOI{10.1145/3411764.3445674}
\acmISBN{978-1-4503-8096-6/21/05}

\usepackage{graphicx}
\usepackage{float} 
\usepackage{subfigure}
\usepackage{multirow}
\usepackage{diagbox}
\usepackage{comment}
\usepackage{enumerate}
\usepackage{xcolor}
\usepackage{ulem}

\begin{document}

\title{Modeling and Leveraging Analytic Focus During Exploratory Visual Analysis}

\author{Zhilan Zhou}
\affiliation{
  \institution{University of North Carolina at Chapel Hill}
  \country{USA}
}
\email{zzl@cs.unc.edu}

\author{Ximing Wen}
\affiliation{
  \institution{University of North Carolina at Chapel Hill}
  \country{USA}
}
\email{ximing@live.unc.edu}

\author{Yue Wang}
\affiliation{
  \institution{University of North Carolina at Chapel Hill}
  \country{USA}
}
\email{wangyue@email.unc.edu}

\author{David Gotz}
\affiliation{
  \institution{University of North Carolina at Chapel Hill}
  \country{USA}
}
\email{gotz@unc.edu}


\pdfinfo{
  /Title (Modeling and Leveraging Analytic Focus During Exploratory Visual Analysis)
  /Author (Zhou, Wen, Wang, and Gotz)
  /Keywords (Analytic Focus, Visual Analytics, User Modeling, Insight Provenance)
}
  
\renewcommand{\shortauthors}{Zhou, Wen, Wang, and Gotz}

\begin{abstract}

Visual analytics systems enable highly interactive exploratory data analysis. Across a range of fields, these technologies have been successfully employed to help users learn from complex data. However, these same exploratory visualization techniques make it easy for users to discover spurious findings. This paper proposes new methods to monitor a user's analytic focus during visual analysis of structured datasets and use it to surface relevant articles that contextualize the visualized findings. Motivated by interactive analyses of electronic health data, this paper introduces a formal model of analytic focus, a computational approach to dynamically update the focus model at the time of user interaction, and a prototype application that leverages this model to surface relevant medical publications to users during visual analysis of a large corpus of medical records. Evaluation results with 24 users show that the modeling approach has high levels of accuracy and is able to surface highly relevant medical abstracts.
\end{abstract}



\maketitle

\input{tex/introduction}
\input{tex/related_work}
\input{tex/platform}
\input{tex/analytic_focus2}
\input{tex/leverage_focus}
\input{tex/user_study}

\input{tex/discussion}
\input{tex/conclusion}

\section{Acknowledgments}

The research reported in this article was supported in part by a grant
from the National Science Foundation (\#1704018).

\bibliographystyle{ACM-Reference-Format}
\bibliography{main}

\typeout{get arXiv to do 4 passes: Label(s) may have changed. Rerun}
\end{document}

%% file: tex/introduction.tex
\section{Introduction}
\label{sec:intro}

Visual analytics technologies are designed to enable exploratory analytical thinking via dynamic visual interfaces \cite{thomas_illuminating_2005}.  These capabilities, coupled with the relatively low technical skills required to utilize well-designed visualization-based analysis tools (in contrast to programming-based tools that require knowledge of algorithms and/or specialized languages) have made them attractive to domain experts across a wide range of disciplines \cite{sun_survey_2013}.

The exploratory nature of visual analysis is a key part of its draw because it provides analysts with the ability to quickly discover new patterns and develop hypotheses from complex data.
Yet along with this powerful capability comes a critical risk.  The more users explore by altering views, changing filters, or applying other data transformations, the more likely they are to come across an ``interesting'' pattern which appears to be potentially meaningful.  This is, of course, part of the draw of visual analytics. However, it is also true that if an analyst examines the data for long enough, they are increasingly likely to find false positives that appear valuable---even statistically significant---eventually \cite{malik_high-volume_2016-1,malik_high-volume_2016}.

Motivated by this challenge, we envision a new approach to contextualizing visualized patterns by surfacing relevant articles interactively during a visual analysis. In this vision, the goal is to help users better see how potentially new insights fit within existing knowledge structures.  Are new discoveries consistent with existing knowledge?  Are observed patterns in conflict with previous observations?  Do relevant articles suggest alternative interpretations or follow-up questions to explore?

A key step in this vision is to computationally model a user's ever-changing analytic focus during exploratory visual analysis.  If an accurate model can be obtained, it can be used to query for relevant documents which can in turn be surfaced via the system's user interface.  

For example, consider the medical domain, a common application area for visual analytics methods \cite{gotz_data-driven_2016}. Imagine a medical expert analyzing a set of medical records for a cohort of heart failure patients. Patterns in treatment that associate with worse outcomes may suggest to the researcher that certain types of care plans are problematic.  However, are these patterns consistent with the medical literature?  As the researcher explores alternative risk factors within the medical record data, are their articles that provide context that would improve their interpretation of the visualized statistics?  Can the literature suggest alternative explanations that motivate the analyst to look elsewhere in the dataset?

A visual analysis system that could automatically model an analyst's analytic focus and surface relevant articles has the potential to help in all of these ways.  However, this vision depends upon the ability to accurately model analytic focus.  This is a challenging endeavor, however, as users do not typically express an explicit definition of their focus.  Instead, the focus of a user's analysis is most often only implicitly expressed through a user's interaction behavior.

This paper describes a set of first steps toward realizing this goal, including a general framework for analytic focus modeling, a prototype implementation, and results from a user study to evaluate our approach.  More specifically, the major contributions presented in this paper include:

\begin{itemize}
\item {\bf Analytic Focus Model and Associated Algorithms.} A formal model designed to represent a user's analytic focus is proposed along with the algorithms required to build and update the focus model over time as a user conducts an exploratory visual analysis.

\item {\bf Prototype Application for Medical Data Analysis.} The proposed approach has been prototyped within a pre-existing visual analysis system~\cite{borland_selection_2020,gotz_visual_2020} designed to discover longitudinal patterns in large collections of structured electronic medical record data.  The prototype implementation leverages this focus model capability to regularly search PubMed abstracts for documents relevant to the user's unfolding analysis.  The relevant abstracts (i.e., relevant to the current focus model) are displayed to users to contextualize the current visualization and to suggest new opportunities for future exploratory analysis.

\item {\bf Evaluation via Controlled User Study.} Results are reported from a 24-person user study conducted to evaluate the proposed approach.  The qualitative and quantitative results evaluate the accuracy of the computed focus model in comparison to manually logged reports of analytic focus during an exploratory analysis of real-world medical data.  The evaluation also examines the utility of the evolving set of relevant abstracts surfaced during the users' analyses.  
\end{itemize}

The remainder of this paper provides an overview of related work and presents a detailed description of the above-mentioned research contributions.  A discussion about the limitations and potential benefits of this type of focus-based contextualization approaches for future interactive systems is also provided.

%% file: tex/related_work.tex
\section{Related Work}
\label{sec:related}

We propose computational approaches for modeling a user's focus during exploratory visual analysis, and leveraging that model to search for contextually relevant literature.  This approach
draws on prior work from various areas in HCI research.

\subsection{Contextual Visualization}

While visual analytics techniques are effective in helping analysts glean insights from large complex data sets, visualized results are often highlighted out of their contexts (e.g. data selection process) which are critical to the validity of analysis. It is therefore important to make analysts aware of such contextual information \cite{borland_contextual_2018}. Techniques to bring back unseen context include showing zoom-out views for spatial, temporal, and network visualizations \cite{morrow2019periphery,bourqui2008revealing}, visualizing data provenance  \cite{gotz2017adaptive}, surfacing relevant information within the same document to the current reading focus \cite{badam2018elastic}, and estimating potential cognitive biases during analytic processes \cite{verbeiren2014pragmatic}. 

Our work differs from previous works in that it brings relevant information from \emph{external} text collection to contextualize structured data visualization.
It saves analysts' effort in searching for published evidence that may support or contradict the current finding. Besides critical thinking, surfacing contextual information from external data can also provide new perspectives and inspire follow-up analyses. While the new methods presented in this paper provide a general framework for analytic focus modeling (see Section~\ref{sec:limitations}), they have been prototyped as new additions to a pre-existing visual analytics system called Cadence~\cite{  gotz_adaptive_2016,gotz2017adaptive,gotz_visual_2020,borland_selection_2020,borland_selection_2021} which includes some contextual visualization features such as selection bias detection and mitigation.  Because of this connection between prior work on Cadence and the new methods presented in this paper, a more detailed description of Cadence (both the pre-existing system and the new additions implemented for this paper) is provided in Section~\ref{sec:platform}.

\subsection{Analysis of User Interactions and Visualization Provenance}

Another related area of research is the analysis of user interactions and visualization provenance. Xu et al.'s recent survey~\cite{xu_survey_2020} characterized this diverse body of work along three dimensions: WHY, WHAT, and HOW. 

The work presented in this paper most directly fits within the ``Understanding the User'' category within the WHY dimension. As defined by Xu et al., this category includes: (1) methods that aim to ``describe the human analytical reasoning proces'' (e.g.,~\cite{dou_recovering_2009,gotz2009characterizing}); (2) computational approaches to extract analysis patterns and/or workflows (e.g.,~\cite{feng_patterns_2019,liu_patterns_2017}); (3) modeling methods to understand user characteristics or personality traits (e.g.,~\cite{brown_finding_2014, ottley_personality_2015}); and (4) techniques for the modeling of user attention and/or tracking biases during analysis (e.g.,~\cite{ottley_follow_2019, wall_toward_2019, wall_warning_2017, gotz_adaptive_2016}).

Within these subcategories, the research presented in this paper is perhaps most closely aligned with work on modeling of user attention. Past work in this area
has often aimed to model \emph{visual} attention (e.g., predicting which visual mark on the computer screen a user will next attend).  This approach, for example, was followed in the attention inference method proposed by Ottley et al.~\cite{ottley_follow_2019} which used a hidden Markov model to predict which mark within a visualization a user will interact with next.  This prediction is based on a mark space defined as a set of $N$ visual features used within a visualization (e.g., position or color of a mark).
The analytic focus model presented here is different. Rather than model and predict the visual marks to which a user attends, this paper aims to model \emph{cognitive} attention: the set of semantic concepts that define the focus of the user's analytic task independent of how or where they are represented visually in the interface. This is modeled by observing semantic user actions (defined using the type of user interaction and associated parameters, see Section~\ref{sec:prob_form}).  In this way, the approach in this paper also has some similarities to Xu et al.'s ``describe the human analytical reasoning process'' subcategory. However, the goal of this paper is to model a user's current analytic focus rather than understanding analytic strategies or identifying findings from interaction logs.

This paper's contributions also relate to the ``Adaptive Systems'' category within the WHY dimension.  In particular, the analytic focus model is designed to be leveraged in ways that adaptively surface relevant information to users during analysis. While formulated differently, this approach has some similarity to past work exploring was to use observations of user behavior to guide data pre-fetching (e.g.,~Battle~et al.~\cite{battle_dynamic_2016}) or to drive active search during visual analysis (e.g.,~Monadjemi et al.~\cite{monadjemi2020active}).  In this characterization, the ActiveVA approach proposed by Monadjemi et al. is perhaps most closely related given its focus on surfacing new data to users that are marked as relevant given an actively updated model of a user's latent interests. However, ActiveVA builds this model from data points that are assigned a binary classification of either relevant or irrelevant based on user input.  In contrast, this paper maintains a semantic concept-based approach with a time-decay model that is maintained by observing properties of user actions (rather than labels of data points).

In the WHAT dimension, the approach in this paper fits within the ``Sequence'' category of Xu et al.'s framework because it observes a sequence of high-level user interactions during visual analysis. One challenge in the analysis of sequences of user interactions is the need to bridge the semantic gap between high-level user intents and low-level user interface events, which Ragan et al. call \emph{granularity}~\cite{ragan2015characterizing} and Xu et al. refer to as \emph{interaction type}~\cite{xu_survey_2020}. The approach outlined in this paper observes user interactions at the semantically meaningful action level as defined in~\cite{gotz2009characterizing}.

Finally, Xu et al.'s HOW dimension characterizes the modeling approach used by a given method.  The methods in this paper are most closely related to the ``probabilistic models/prediction'' category.  This is a broad category ranging from basic statistical models to more sophisticated prediction methods including neural networks (e.g.,~\cite{guo_visualizing_2019, smith_predicting_2018}) and Markov models (e.g.,~\cite{ottley_follow_2019, wei_visual_2012}).  The method proposed in this paper uses a time-decay model to computer per-concept importance scores as predictions of salience to a user's analytic focus at a given time point.

With respect to implementation, the methods presented and evaluated in this paper have been prototyped as new components within a pre-existing visual analytics system (see Section~\ref{sec:platform}). However, past work on VisTrails~\cite{bavoil2005vistrails} and the recently published Trrack \cite{2020_visshort_trrack} show that a library-based design can be used to support generic provenance tracking capabilities.  Within the scope described in Section~\ref{sec:limitations}, a similar approach could be used to apply the methods presented in this paper beyond our prototype environment.

\subsection{User Modeling \& Recommendation}

User modeling is another closely related active research area in human-computer interaction~\cite{fischer2001user,biswas2017wiley}. The goal of user modeling is to enable interactive systems to better understand users' intent and preferences so as to provide customized support to satisfy users' specific needs.  User modeling techniques often infer user preference from user behavior logs on interface components. The technique has been widely adopted in Web search engines \cite{white2016interactions}, digital libraries \cite{frias2006automated}, online recommender systems \cite{berkovsky2008mediation}, adaptive education systems \cite{brusilovsky2007user}, and health information systems \cite{zheng2015computational}. In this work, we model a user's focus by observing user interactions on structured data elements and representing user interests using domain-specific concepts. Such an approach is closely related to ontology-based user modeling, where users interact with the semantic web and user profiles are mapped to structural elements in an ontology  \cite{sosnovsky2010ontological}.

%% file: tex/platform.tex
\section{Visual Analysis Platform}
\label{sec:platform}

This paper introduces a set of techniques designed to model and leverage a user's analytic focus during exploratory visual analysis.  These techniques, presented in detail in Section~\ref{sec:analytic_focus}, are designed as a general approach that can be tailored to work across a broad range of visual analysis applications.  However, the design is motivated in part by challenges faced by analysts in the medical domain and the techniques have been prototyped for evaluation purposes within Cadence~\cite{gotz_adaptive_2016,gotz2017adaptive,gotz_visual_2020,borland_selection_2020,borland_selection_2021}, a pre-existing open-source \cite{cadence_github} visual analytic system designed to enable exploratory analysis of large collections of longitudinal data such as electronic health records. This section provides a brief overview of the Cadence system and the user interface extensions added to support the work presented in this paper.

\subsection{Pre-existing Cadence System}

Cadence is a visual analytics platform designed to allow users to discover patterns in structured longitudinal event data, such as electronic health records.  Cadence provides a rich set of interactive features for defining queries, applying filters, exploring the frequency of different patterns of events over time, and associating those event patterns with differences in outcome.  For example, Cadence has been used to analyze large collections of medical data to discover risk factors for opiate addiction, identify treatment pathways for heart failure patients, and discover patterns of use for medical devices.

Cadence employs a generic model to represent longitudinal event data. It consists of a subject information table and an event sequence collection. The subject information table contains a $subject\_id$ column as the primary key and various named attributes that describe the subject. The event sequence collection stores $(eventtype, timestamp)$ pairs for each subject. Cadence understands an $eventtype$ as a nominal value in a fixed vocabulary, and organizes the types within a hierarchical structure (e.g. ``is-a'' relation) to support aggregation of events at different levels of granularity. A $timestamp$ records the date and time of $event$. 
In the context of electronic health records, the subject table contains patient attributes (such as age, gender, and race). The event sequence collection stores medical events over time for each patient, where the event types are concepts (such as procedures and diagnoses) from standard medical lexicons such as SNOMED-CT and ICD-10.

\begin{figure}[t]
\centering  
\includegraphics[width=0.85\textwidth]{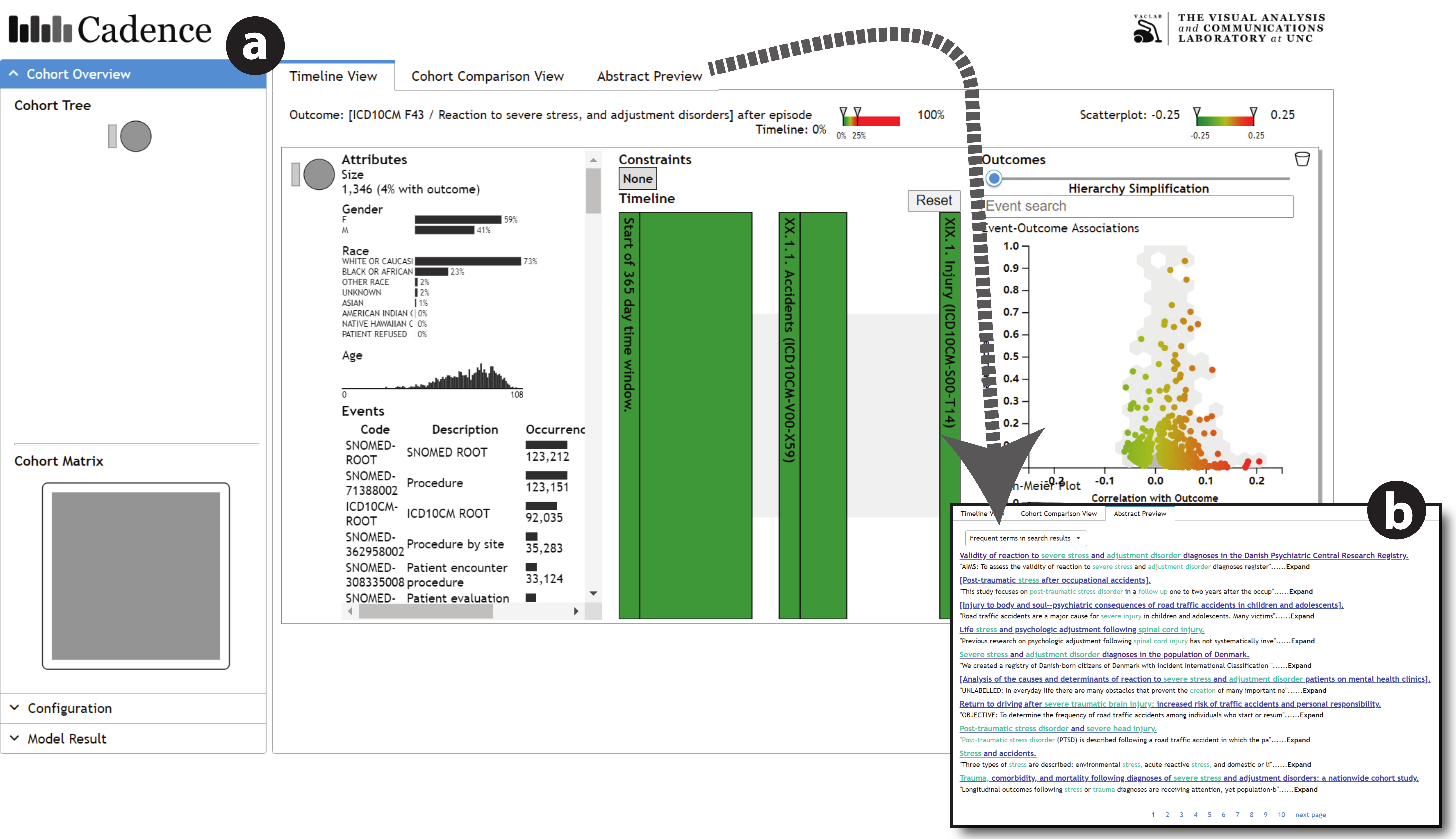}
\Description[Two UI screenshots of Cadence]{An UI screenshot containing interactive histograms, timeline, and a scatterplot and another screenshot containing a list of abstracts}
\vspace{-0.2cm}
\caption{Cadence is a visual analytics platform that allows for exploratory analysis of longitudinal event data such as electronic health records.  (a) This pre-existing system contains a number of visualization features including interactive histograms, an exploratory timeline, and a highly interactive scatterplot showing associations between specific event types and outcomes (e.g., specific treatment events associated with hospital discharge or death). (b) Cadence has been extended to surface relevant abstracts from PubMed using the focus model techniques presented in this paper.}
\label{fig:cadence}
\end{figure}

The Cadence platform provides a web-based interface shown in Figure~\ref{fig:cadence}(a) which includes a coordinated set of interactive visualizations that work together to allow users to explore complex event data over time. 
A user starts her/his analytic task by defining a set of constraints as a scoping query.  The query is defined via interaction with a drag-and-drop query authoring tool as a set of temporal and attribute constraints. For example, in the medical context, a health analyst might query a database of millions of medical records for data from patients that are male, over age 65, with a diagnosis of heart failure, and with a hospital discharge later. To conduct an analysis about readmission risk factors, the analyst might specify the temporal query to include all medical events for matching patients occurring from one year prior to heart failure diagnosis through 90 days after the hospital discharge. Given this type of query specification, the platform then selects the data satisfying the query, computes various summary statistics in the backend, and renders the results as interactive visualizations in the frontend as dynamically generated interactive visualizations using standard Javascript libraries (e.g. D3.js). The user interface for Cadence with some of the available visualizations is shown in Figure~\ref{fig:cadence}.  We note that this figure does not show all of the available visualizations within Cadence as only a subset of visual elements is displayed at any one time to manage interaction complexity.
Moreover, we note that this figure shows new user interface elements that were not part of the pre-existing Cadence system but instead as part of the work reported in this paper as described in Section~\ref{sec:platform_new}.  In particular, the Model Result component (on the bottom left) and the Abstract Preview panel (highlighted in Figure~\ref{fig:cadence}(b)) were not part of the pre-existing Cadence system.

The pre-existing Cadence system allows users to visually explore the data returned by the query and to apply filters via the interface to revise the subset of event data under analysis.  A record of data subsets (which Cadence calls cohorts created by users is maintained and visualized through the Cohort Overview panel found in the left sidebar of the interface.  This panel shows iconic representations of the cohorts that summarize basic information (such as how the subsets were derived, and the number of entities in each subset).  In addition, Cadence uses this area to surface bias statistics calculated by system features that quantify potential selection bias and support selection bias mitigation.  These features are beyond the scope of this paper and are reported in prior work \cite{borland_selection_2020,borland_selection_2021}.

Within the main portion of the pre-existing Cadence user interface (labeled as ``Timeline View'' in Figure~\ref{fig:cadence}), a number of coordinated visualizations are provided to explore the data associated with a specific cohort and to apply filters to derive new cohorts. First, interactive histograms allow users to subset data based on categorical or scalar attributes such as age, race, or gender. Users can right click on any of the histograms to apply a filter.  In response, the system derives a new cohort by applying the requested filter and updates the visualization to reflect the updated data subset.  Below the attribute histograms, a histogram of event frequencies is provided to communicate the most commonly occurring events within the currently visualized data subset.  Leveraging the fact that event types can be organized within a hierarchy (as described earlier in this section), this portion of the user interface includes data at various levels of granularity.  Moreover, Cadence includes a variety of features that allow users to interactively control the level of aggregation used in the interface to analyze events at various levels of detail (e.g., to look at all ``heart disease'' diagnoses as one type of event, or to look 19 distinct types of heart failure diagnoses at the lowest level of representation).  These features are valuable analytical tools for managing high-dimensionality, but are beyond the scope of this paper and reported in prior work \cite{gotz_visual_2020}.

To the right of the histograms is a milestone-based timeline visualization~\cite{gotz_decisionflow:_2014} which allows users to explore subsets of data based on temporal constructs between events (e.g., combinations of before/after relationships).  The rectangular blocks in the visualization correspond to specific types of events (e.g., a diagnosis of some sort of injury due to an accident) or temporal milestones (e.g., 365 days prior to the subsequent event) of the event sequences being analyzed. Users can click on individual rectangles to view details-on-demand about the corresponding event sequences.  Most attribute and temporal constraints in the scoping query are shown as the initial milestones.  Moreover, users can interactively add or remove milestones within the timeline to create further subdivisions (represented visually via the insertion of additional rectangles) via interactions with the scatter plot described later in this section.  The rectangles themselves are color-coded to represent outcomes.  In the medical context, this allows users, for example, to example different clinical pathways (e.g., different treatments for a common condition) and the associated medical outcomes experienced by the corresponding patients.  If desired, users can derive cohorts from individual rectangles to narrow their analysis to a specific subset.  This filtering action is triggered via a right-click context menu, and it results in the creation of a new cohort via the introduction of a new data constraint just like a filtering action performed via the attribute histograms described earlier.

Selections within the timeline visualization are coordinated with a scatter-plus-focus plot~\cite{gotz_visual_2020} which allows users to quickly identify new event types to incorporate into the timeline to help separate patients with good outcomes from those with bad outcomes. The plot shows circles that represent individual event types positioned by frequency on the y axis and correlation to the outcome on the x axis. Users can click on individual circles to view more details about individual event types, navigate the event type hierarchy, and add interesting events found in this visualization to the timeline visualization as new milestones.

As this description of the pre-existing Cadence system shows, users can perform various actions through the user interface during exploratory analysis. This includes the addition of new filters, inserting milestone event types into the timeline view, or clicking on graphical marks that represent subgroups (defined by combinations of milestones in the timeline) or event types for details-on-demand.  The Cadence system, in response to these actions, computes and visualizes a number of different types of statistics for the various data subsets created by users during their exploratory analyses.

The collection of data subsets recorded by the pre-existing Cadence system (representing groups of patients in the medical context) is used to allow users to go back to data sets viewed in prior stages of analysis, and 
to support the selection bias detection features described above. This is accomplished by comparing the users' current data subset against a baseline data subset seen earlier in an analysis. Users can interactively select past data subsets as baselines for this purpose through interactions with the Cohort Overview panel. The pre-existing Cadence system had no capabilities for monitoring user actions during exploratory analysis, constructing and maintaining a model of user's analytic focus during analysis, or leveraging such a focus model in any way.  Moreover, no user interface components existed to support these features which were introduced to Cadence for the work presented in this paper as described below.

\subsection{New Additions to Cadence to Support the New Focus Model Capability}
\label{sec:platform_new}

Following the new approach we outline in this paper (and therefore not part of the pre-existing Cadence system), we can in theory describe each of user analysis actions supported by Cadence as a combination of both (1) a type of action and (2) one or more event types (i.e., concepts) that are the focus of that action. For instance, a user might add a new filter (the type of action) using a diagnosis of heart failure (the concept of that action) in medical analysis to focus only on patients with that diagnosis.  In this way, we are able to apply the focus modeling approach presented in this paper (as formulated in Section~\ref{sec:prob_form}) to the pre-existing Cadence visual analytics system.

To adopt this model of user actions and to implement the new focus model capabilities outlined in this paper, a wide range of new additions were made to the pre-existing Cadence system. In particular, to support the new focus model capabilities, the techniques described in Sections~\ref{sec:analytic_focus} and~\ref{sec:leveraging_focus} have been prototyped as extensions to the pre-existing Cadence visual analytics system. The added features automatically observe a user's analytic activity during exploratory analysis of the structured medical data, incorporate those observed interactions into a model representing the user's analytic focus as it evolves during analysis, and leverages that model to asynchronously retrieve relevant abstracts from PubMed with the aim of contextualizing the temporal event patterns surfaced in the pre-existing visualizations. 

To support these added capabilities, a new panel has been added to the Cadence user interface as shown in Figure~\ref{fig:cadence}(b).  This is visible as an additional tab located behind the timeline visualization.  The text snippets shown for each abstract include highlights of any mentioned medical concepts that are related to the current focus model.  Users can switch between tabs at any time to look for linkages between the medical literature and patterns found in the structured data visualization.

Beyond the core extensions outlined above, which were designed to assist users during analysis, an additional focus model panel was added to the interface in support of the user study protocol outlined in Section~\ref{sec:study}.  This added panel is not intended for everyday use.  Instead, it externalizes the current focus model for user review during the study protocol, provides buttons that allow for study moderators to manually save data during a study session (such as logs of users' interactive analysis activity and the current state of the focus model) to allow for subsequent analysis. 

%% file: tex/analytic_focus2.tex
\section{Analytic Focus Modeling}
\label{sec:analytic_focus}

When visually exploring structured data, a user aims to find interesting patterns expressed through domain-specific concepts and their relations. 
Because the task is intrinsically exploratory, the user's analytic focus will change over time. Some of these concepts come into the user's current focus, while others fade away. The problem of analytic focus modeling can be stated as ``to infer a user's current analytic focus given the trace of user actions observed on the system interface.'' 
In this section, we formulate this problem and propose an algorithm that estimates the focus model over time.

\begin{figure}[ht]
\centering  
\includegraphics[width=0.5\textwidth]{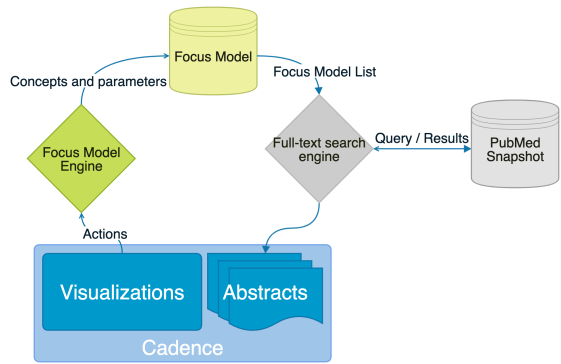}
\Description[A flowchart]{A flowchart depicts the circulatory workflow from the Cadence visualizations to the Focus Model Engine and the Focus Model, then to the full-text search engine queries and receives results from the PubMed snapshot, finally to the Abstracts in Cadence}
\vspace{-0.6cm}
\caption{As users interact with the visualizations, actions are reported to the Focus Model Engine which updates the focus model. The full-text search engine converts the Focus Model into a query that is used to retrieve relevant PubMed abstracts.}
\label{fig:flowchart}
\end{figure}

\subsection{Problem Formulation}
\label{sec:prob_form}

Before formulating the problem of analytic focus modeling, we first define several key terms and notations.

\textbf{Definition 1 (Interaction)}: An \emph{interaction} refers to low-level UI events, such as clicks, drags, or key-presses. Each interaction usually carries little semantic meaning by itself. A group of interactions (e.g. a right-click and then a click on an option in the context menu) can accomplish higher-level semantics, which we define below.

\textbf{Definition 2 (Action)}: An \emph{action} refers to an atomic semantic operation at the level where a series of aforementioned \emph{interactions} combine to represent a single semantically meaningful event like a selection or a filter. Each action provides event-based insight provenance that carries richer semantics than an interaction. 
An action can be represented as a tuple combining type, intent and parameters. \cite{gotz2009characterizing}  The type could be a query, a selection, or a filter.
Our system does not use intent, resulting in a simplified action representation of (type, parameters). We instrument the visual analytics system such that it will listen to a selected set of actions taken by a user.  The set of all actions $A$ is described in the first column of Table \ref{tab:action_cadence}. In what follows, we use $a \in A$ to denote an action.

\textbf{Definition 3 (Concept)}: A \emph{concept} refers to a meaningful data attribute in the problem domain. These can be attributes (fields) of data records in a structured database schema, index terms appearing in unstructured documents, and entries in a domain-specific taxonomy. 
Concepts are action parameters in the action tuple, and vary by the type. For example, selecting is usually associated with at least one concept - the selection criteria.
Concepts play a central role in a user's analytic focus. We use $c$ to denote a generic concept.

\textbf{Definition 4 (Persistent Action)}: A \emph{persistent action} changes the visualization interface significantly in the long run, until other persistent actions are performed and revert the change. For example, adding a filter in most visualization tools changes the plot persistently until the user removes the filter or adds another filter.

\textbf{Definition 5 (Transient Action)}: A \emph{transient action} changes the visualization interface and those changes revert back quickly, and sometimes automatically. For example, the user could select a point in a scatter plot by hovering to check the details of the data it represents, but these details will fade out after 5 seconds even the cursor is still hovering and the user does not interact with any other parts. 

We categorize all actions in $A$ into the above two types. The second column of Table~\ref{tab:action_cadence} shows the type for each action.

\textbf{Definition 6 (Time Step)}: During exploratory data analysis, the user's analytic focus will change over time. Although wall clock time is a natural way to define time, different users may perform the analysis at different paces, depending on the path of exploration and the user's familiarity with the domain. In light of this, we measure time as the sequential order of actions. That is, each \emph{time step} $t$ is a sequence number that corresponds to a discrete action taken anywhere on the interface. In particular, $t = 0$ corresponds to the first action, the initial query issued by the user.

Using the above definitions, we can define the observed sequence of user actions throughout the data exploration process as $S = \{(a_i, t_i, c_i) | 0 \le t_i \le T\}$, where each triple $(a_i, t_i, c_i)$ represents an action $a_i\in A$ taken place at time step $t_i$ that involves a concept $c_i$. $T$ is the total number of time steps (actions) taken. In principle, the index $i$ can be different from the time step $t_i$. For instance, an action may simultaneously involve two concepts, which gives rise to two triples, each with the same action, the same time step, but two different concepts. In that case,  $t_i$ will only increment by 1 while $i$ will increment by 2.

\textbf{Problem Formulation (Analytic Focus Modeling)}:
At any time $t \le T$, $S_t = \{(a_i, t_i, c_i) | 0 \le t_i \le t\}$ will contain a unique set of concepts $C_t$. The \emph{Analytic Focus Modeling} problem is as follows: given the observed sequence of actions $S_t$, to estimate an importance score $I^c (t) \ge 0$ for each concept $c \in C_t$ at time $t$. A large importance score $I^c (t)$ indicates a strong analytic focus on concept $c$ at time $t$; a small importance score indicates a weak focus; a zero importance score indicates the concept falls out of  focus.

\begin{table}[ht]
\small
\begin{tabular}{@{}lrrrr@{}}
\toprule
Action $a$                &  Category     & $I_a(0)$ & $P_a$ & $b_a$ \\ \midrule
Select (Scatterplot) &  T      & 1        & 2    & 0   \\
Select (Timeline)    &  T      & 1        & 2    & 0   \\
Add milestone       &  P     & 1        & 50   & 0   \\
Add filter          &  P     & 2        & 80   & 1 \\
Query               &  P     & 3        & 100  & 2   \\
Close (Panel)       &  P     & -0.1     & $+\infty$   & 0  \\
Reset (Panel)       &  P     & -0.3     & $+\infty$   & 0  \\
Show timeline       &  P     & 1        & 80   & 0   \\
Set baseline        &  P     & 1        & 100  & 0   \\
Set focus           &  P     & 1.5      & 100  & 0.5   \\ \bottomrule
\end{tabular}
\Description[A table of scores with corresponding action types]{A table of importance scores, persistence scores and the importance bias scores with corresponding action types and categories}
\caption{User actions considered in this study, with associated parameters. 
P stands for persistent; T stands for transient.
$I_a(0)$, $P_a$, and $b_a$ are the initial importance score, the persistence score, and the importance bias score of action $a$, respectively.
}
\label{tab:action_cadence}
\end{table}

\subsection{Analytic Focus Modeling Algorithm}

Given the above formulation, we describe an algorithm that estimates a dynamic focus model given a user's actions.

\subsubsection{An Additive Model}
Each concept $c$ in the focus model is associated with an \emph{importance score} $I^c(t)$. To estimate $I^c(t)$, we take all  actions related to concept $c$ up to time $t$: $S_{t,c} = \{(a_i, t_i)|0\le t_i \le t\}$. In general, these may include not just actions that directly involve $c$ as an input, but also actions adjacent to concept $c$ on the interface. For example, if an action involving a different concept took place in the same panel as concept $c$, that action is related to $c$.

We estimate $I^c(t)$ by aggregating partial importance scores contributed by all previous actions involving concept $c$:
\begin{align}
I^c(t) = \sum_{(a_i, t_i) \in S_{t,c}} I_{a_i} (t - t_i) \ ,
\end{align}
where $I_a(t)$ represents the importance of a concept as a result of taking action $a$ on it. For example, creating a filter using a concept indicates that the concept is in the user's current focus and will remain there for some time. Note that  $I_a(t)$  is action-specific but concept-agnostic. This is a simplifying assumption that an action will contribute the same importance regardless of concepts involved. $I_{a_i}(t-t_i)$ is a time-shifted version of $I_{a_i}(t)$ because the importance contribution starts from $t_i$, the time when action $a_i$ happened.

In principle, it is possible to use other strategies to aggregate per-action importance scores and produce the overall importance score $I^c(t)$, e.g., taking maximum instead of sum. We adopt an additive model in this preliminary study.

\subsubsection{Per-Action Importance Score Function}

For each action $a_i$ happened at time step $t_i$, we associate a decaying \emph{aging function} over time. The basic idea is simple: as time goes by, the importance of concepts involved in that action will decline. In particular, we employ the approximate Ebbinghaus forgetting curve~\cite{wozniak1995two} as our aging function:
\begin{align}
I_a(t) = I_a(0)\times e^{-\frac{t}{P_a}} \ , 
\end{align}
where $P_a$ refers to the Persistence Score of action $a$ and $t$ refers to the time step counter indicating the number of actions after $a$ took place. We utilize the Ebbinghaus forgetting curve as the aging function because the focus on a certain concept based on one action shares similarities with the retention of facts in memory. Our Persistence Score represents the stability of the action influence, working similarly like the $S$ as stability of memory in the original Ebbinghaus curve equation $R = e^{-\frac{t}{S}}$. $I_a$ will decrease fast at the beginning and then slowly decrease. When it is below some certain lower threshold $l$, the Focus Model will remove this aging function in the list to improve efficiency.

When $t = 0$ for an aging function, the initial value $I_{a}(0)$ is assigned based on the type of action $a$. 
A larger number for an action represents the semantic importance of that action in terms of a user's analysis focus. For example, the filter action that changes the dataset in examination is a much stronger signal of analytic focus than a transient inspection action; hence a larger $I_{a}(0)$ will be assigned to the filter action.

The Persistence Score $P_a$ implies how stable an action influences the user interface determined by its type. A higher persistence score implies that the action makes more stable changes on the interface. We use such a standard based on the observation that users tend to modify the visualization interface to help their analysis. So when an action adds certain components on the interface stably, the values corresponding to the components are more important in the user's focus. Thus, the persistent actions are assigned with higher persistence scores than the transient actions. 

Some actions like ``reset'' or ``undo'' will remove components on the interface, they are persistent actions with negative initial importance. We set their persistence score to $+\infty$ as we assume negative effects do not change over time. That is, $I_a(t) = I_a(0)$ is a negative constant function for these actions.

\subsubsection{Updating the Focus Model}

Each time the algorithm receives a new action $(a_i, t_i)$, it needs to update the Focus Model. It mainly adds new aging functions to relative concepts of the action inputs and updates importance scores. This updating process is comprised of two parts: independent decay of each aging function and sum of all aging functions, including the newly added ones for those concepts that are close enough with the action inputs.

We consider the updating process of the overall importance score as an incremental operation. When time is at $t+1$, concept $c$'s importance score is updated as follows:
\begin{align}
I^c(t+1) \leftarrow I^c(t) + \frac{\mathrm{d}I^{c}(t)}{\mathrm{d}t} \Delta t + I_{a} (0) \ ,
\end{align}
where $I_{a} (0)$ is the initial importance score of an action $a$ happened at time $t+1$. Since the time steps are discrete, $\Delta t = 1$.

\subsection{Implementation}\label{sec:fm_cadence}

The focus model system is designed to work generically without any dependence on a specific visual analysis system. It is self-contained with a defined API and works with any arbitrary set of concepts.  To utilize the focus model,
developers would need to perform three steps: (1) enumerate action types and parameters, (2) instrument the visual analysis to report the occurrence of actions (and associated parameters) via the API, and (3) tune action parameters to account for application-specific differences in the salience of specific action types.

Following this approach, we developed a \textbf{Focus Model Engine (FME)} as a JavaScript module. The FME then connects with the previously existing Cadence visual analytic system (described in Section~\ref{sec:platform}) through callbacks. Cadence notifies the FME when user actions occur and in response updates the focus model dynamically. Cadence can then request the current focus model at any time via API to support focus-aware features (e.g., see Section~\ref{sec:leveraging_focus}).

In the context of the Cadence system, we considered 10 unique Cadence actions as shown in Table~\ref{tab:action_cadence}. The concepts are medical terms from both ICD-10 and SNOMED-CT coding systems. Action parameters were assigned based on their action categories and heuristics. We also validated these parameters in pilot runs within the authors.

We use an additional importance bias score $b_a$ (see values in Table~\ref{tab:action_cadence}) in the per-action importance score function $I_a(t)$ to encode the intrinsic importance given by an action to an involved concept:
\begin{align}
I_a(t) = I_a(0) \times e^{-\frac{t}{P_a}} + b_a \ . 
\end{align}

In principle, we expect that different parameters are needed for different visual analysis platforms, since they may contain different actions and similar actions may represent different semantic meanings in specific environments. We refer the reader to previous works on classifications of actions and abstract visualization tasks~\cite{brehmer2013multi,sedig2012towards,von2014interaction,schulz2013design}.

%% file: tex/leverage_focus.tex
\section{Leveraging Analytic Focus}
\label{sec:leveraging_focus}

In this work, we leverage the focus model as a succinct representation of the user's information need and use it to retrieve relevant documents from a large text collection. Specifically, we implement a full-text search engine.
A PubMed snapshot from late 2019 is used as our collection of medical articles. It contains  29,137,784 published articles, including each article's title, abstract, authors, subject headings, among other information. We use Apache Lucene/Solr to build a full-text index for the title and abstract of each article. Clinical concepts mentioned in an article's title and abstract are recognized using the SNOMED-CT lexicon.

An analytic focus model is translated into a fielded free-text search query run against the above Solr index. Specifically, the list of (concept, importance score) pairs in the focus model is transformed into a Lucene query. Concept descriptions are used as query terms, and the associated importance scores are used as per-term boost scores defined in Lucene's query syntax. As a preprocessing step, ontology codes (such as ``F43'', ``V00-X59'') and stop words (such as ``of'', ``to'') in concept descriptions are removed from the query. Both title and abstract fields are searched.
For example, given the list of importance-weighted concepts as follows:

\vspace{.05in}
{\small
\noindent
\texttt{[(F43 Reaction to severe stress, 7.3), (V00-X59 Accident, 4.5), (S00-T14 Injury, 2.5)]}
}
\vspace{.05in}

The list is translated into the following Lucene query, which is sent to the Solr search service:

\vspace{.05in}
{\small
\noindent
\texttt{title:reaction\^{}7.3 abstract:reaction\^{}7.3 title:severe\^{}7.3 abstract:severe\^{}7.3 title:stress\^{}7.3} $\cdots$

\noindent
$\cdots$ \texttt{abstract:stress\^{}7.3
title:accident\^{}4.5 abstract:accident\^{}4.5 title:injury\^{}2.5 abstract:injury\^{}2.5}
}

\vspace{.05in}
Upon receiving such a query, the Solr index returns a ranked list of relevant articles. Articles are ranked using the BM25 scoring function, which is a standard document retrieval function that estimates the degree of relevance between a document and a query \cite{robertson2009probabilistic}.  100  top-ranked documents are returned to the frontend of Cadence system. 

As shown in Figure \ref{fig:cadence} (b), the search results are rendered as a paginated ranked list in a separate ``Abstract Preview'' tab beside the main visual exploration tab. Clinical concepts mentioned in the result snippets are highlighted. To summarize clinical concepts mentioned in the search results, 10 most frequently mentioned concepts in the result set are shown in a drop-down menu, which can be used to filter search results.

As the focus model keeps updating during exploratory visual analysis, new search queries are continuously formulated and new results are updated in the ``Abstract Preview'' tab. Because all clinical concepts are pre-extracted at index time (instead of search time), the text search and rendering process is fairly responsive. Using a 32GB RAM, 120 GB SSD server in AWS cloud, the Solr index achieves a response time within 1-2 seconds per query.

%% file: tex/user_study.tex
\section{User Study}
\label{sec:study}

We conducted a user study with 24 participants to test the accuracy of our Focus Model and the helpfulness of the related abstracts returned from PubMed. This section describes the study's design, data processing methods, and results.

\subsection{Study Design}

Our study asked participants to independently perform an analysis using Cadence for a specific task that we provided. Participants were trained to use Cadence prior to the experimental task, and data was collected from both the participant and the system while the task was completed. Post-task information was gathered via a questionnaire and semi-structured interviews.  The details of this study design, which we conducted with the approval of our institutional review board, are provided in the remainder of this section.

\subsubsection{Participants}

To ensure that the participants could perform the analysis smoothly, besides the training, we require the participants to have data analysis experience previously. We recruited 24 current or former graduate students whose graduate studies are/were in a STEM field. 
Although we ask participants to analyze medical data, we do not expect them with a professional medical background. Instead, we provide analysis tasks that can be analyzed using common sense.
Each participant is rewarded with a 15 dollar Amazon gift card.

\subsubsection{Procedures}

Since Cadence is a complex platform, we provide the participants with a tutorial video on using Cadence with the EHR data before the session. We also provide background stories to explain the tasks they are going to analyze during the study session. For example, one of the tasks is to explore what diagnoses or procedures have a strong correlation with a final diagnosis of \textit{reaction to severe stress} (ICD-10 code: F43). We set a specific patient cohort who had an \textit{accident} (ICD-10 code: V00-X59) followed by an \textit{injury} (ICD-10 code: S00-T14) to help start the analysis. We also explain further the relationships among these diagnoses to help those without a medical background. We ask the participants to further explore what other diagnoses or procedures may increase the chance of developing a \textit{reaction to severe stress} among these patients.

During the session, Cadence provides participants with access to interactive data visualizations of aggregate statistics from electronic medical data, and a list of abstracts from related PubMed articles. The Cadence system is instrumented to interactions the participants performed in the system, and compute based on those interactions a Focus Model in real-time.

Participants are asked to pause their analysis approximately every 6 min. Since each participant has a different speed to get familiar with the system, the actual time duration varies among participants. During that pause, they are asked to describe their current analysis focus via a Focus Description Questionnaire. We then show them our computed Focus Model and ask them to compare that model to their own description of their analytic focus. They may modify the questionnaire like adding concepts they forget.

At the beginning of the sessions, we encourage the participants to check the Abstract List during their analysis, but that is not required. If the participants do not check the Abstract List, we let them explore it for about 2 minutes before they fill in the post-task questionnaire.

Whether the participants finish the task or not, they are asked to stop working on their analysis after about 40 minutes. We then provide participants with a post-task questionnaire and conduct a semi-structured interview to collect additional subjective feedback on the Focus Model and related abstracts.

\subsection{Results}

In this subsection, we present our study results in three parts: the comparison between participants' self-reported concept list and our Focus Model result; the aggregated data collected from the post-task questionnaire; and the common answers from the interview.

\subsubsection{Computed accuracy of the Focus Model}
To compute the accuracy, we calculate the recall of our Focus Model, the Jaccard distance and the edited distance between the self-reported concept list and the Focus Model result for each participant and each pause. In our Focus Description Questionnaire, we ask participants to report any concepts which are important in their focus, thus participants may provide certain medical terms within the SNOMED or ICD-10 system, or other related terms.
Hence we need to standardize the questionnaires and the Focus Model results first before the comparison.
We use both the participants' lists before and after their modification during each pause and compare them separately.

First, we need to standardize the concepts in the participant's descriptions of their focus and find the mappings in our Focus Model result.
For concepts that are just a shorthand of a specific code, we directly map the concept to the corresponding code: if a participant wrote \textit{"lead ECG"} in the focus description, we map it to \textit{"268400002: 12 lead ECG"}. However, participants may provide general medical terms rather than a specific diagnosis like \textit{obesity}. For such general concepts, we need to map them to the codes used in our focus model. 
We adopt the following two rules when we do the mapping:

\begin{itemize}
\item{\textbf{Mapping down in the hierarchy.}} We only map a general concept to its child concepts in ICD-10 or SNOMED but we don't do the opposite. For example, if a participant mentions \textit{"procedure"}, we map this concept to the code \textit{"59108006: Injection"} in Focus Model result given the SNOMED categorization. However, if another participant mentions "eye injury", we don't map the concept to code \textit{"S00-T14: XIX.1. Injury"} as the \textit{eye injury} is a child to \textit{Injury}.

\item{\textbf{Combining codes.}} We map a concept to one or multiple codes if all the codes are proper. For example, if a participant mentioned \textit{"long term drug use"} in the focus description, and we have \textit{"Z79.82: Long term (current) use of aspirin"}, \textit{"Z79.84: Long term (current) use of oral hypoglycemic drugs"} and \textit{"Z79.8: Other long term (current) drug therapy"}, we could map this concept to all the three codes. Note that although \textit{Z79.8} is the parent of both \textit{Z79.82} and \textit{Z79.84}, we count them as independent concepts because our prototype also treats these three concepts independently.
\end{itemize}

In some rare cases, participants may also provide concepts like "the outcome scatterplot" which are in their analytic focus but not medical related terms at all. We exclude these concepts in comparison as these concepts do not participate in the actual reasoning process in data analysis.

Using such a mapping strategy, we could always get one-to-one or one-to-multiple mappings from participants' self-reported concept list to our Focus Model result in practice.

Second, we calculate the importance score for each concept after the standardization.
If a concept is mapped to only one code in the Focus Model, we assign the Important Score of that code to this concept. If a concept is mapped to multiple codes in the Focus Model, we assign the sum of the Important Scores of all codes to it.

Third, we compute the ordered list of both concepts described by participants and in the Focus Model result for each pause. For the concept set $V_u$ described by each participant in each pause, we sort all the concepts based on their scores from large to small and get an ordered list $L_u$. 
For the concept set in the Focus Model, we combine those concepts if a one-to-multiple mapping described previously happens.

The new concept set is $V_m$. Then we sort the new concept set based on each concept's score from large to small, and ignore concepts whose scores are below the last concept mapped in the list. Thus we get the final ordered list $L_v$.

We use the following metrics to measure the match between user-provided focus and the focus model:
\begin{enumerate} [(1)]
    \item \textbf{Recall}: the fraction of user-provided concepts covered by the focus model. \ \ Recall $= |V_m \cap V_u| /|V_u|$.
    \item \textbf{Jaccard distance} between two sets $V_u$ and $V_m$.  \ \ Jaccard distance $ = 1 - |V_m \cap V_u| /|V_m \cup V_u|$.
    \item \textbf{Edit distance} between two ordered lists $L_m$ and $L_m$. 
\end{enumerate}

\begin{figure}[ht]
\centering  

\includegraphics[width=1\textwidth]{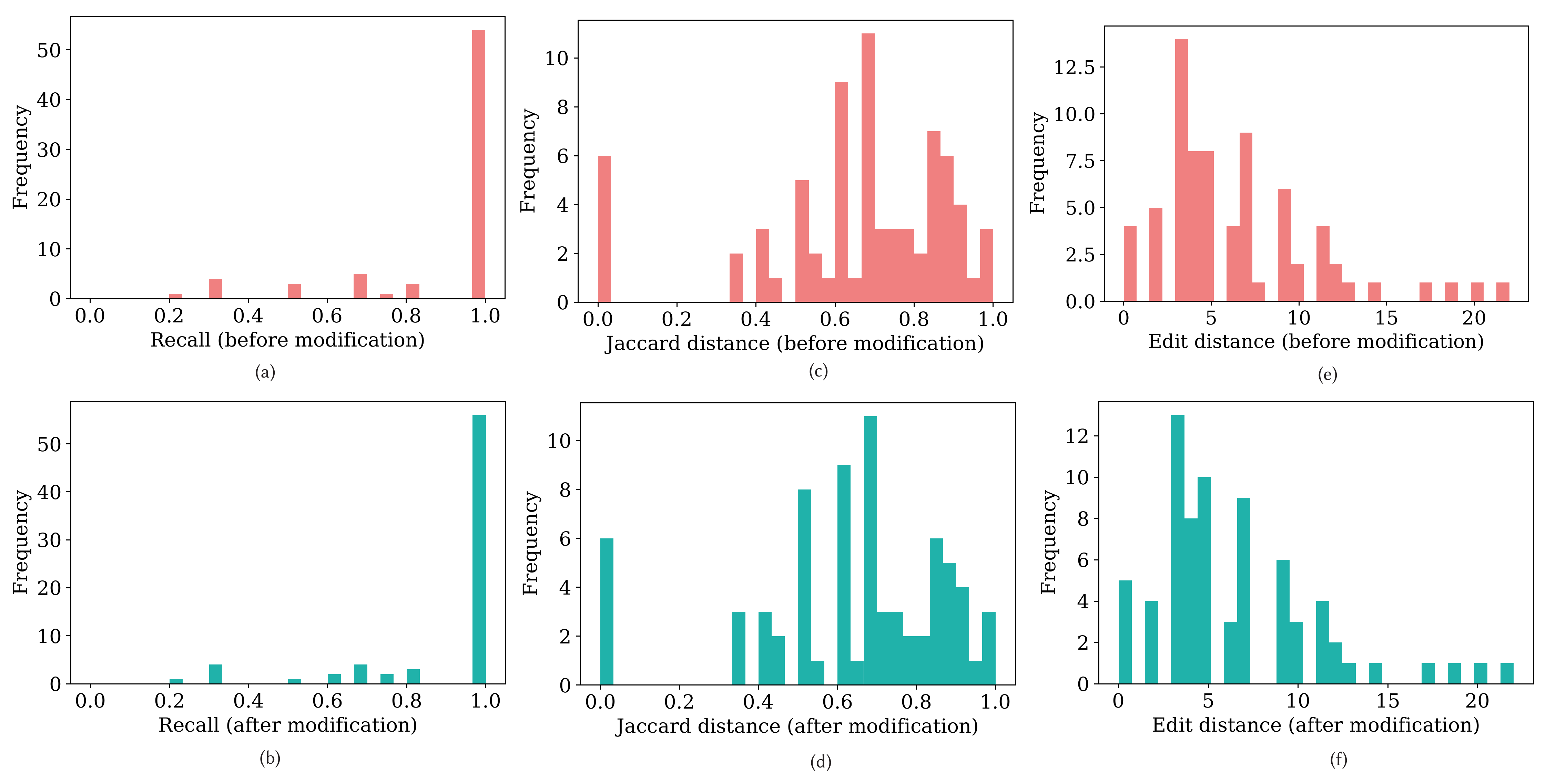}
\Description[Six Histograms]{Six histograms showing the distributions of recall, Jaccard distance and edit distance with two colors indicating before or after the modification}
\caption{Distributions of recall, Jaccard distance, and edit distance between user-provided focus and system-estimated focus model. Participants are allowed to modify their self-reported concepts list after checking our Focus Model result. We separate the scores before and after the modification if it happens. Recall: the larger the better; Jaccard distance and edit distance: the smaller the better.}
\label{fig:concepts_distribution}
\end{figure}

The distribution of these three measurements are shown in Figure~\ref{fig:concepts_distribution} and Table~\ref{tab:mean_medium} for both results before or after the modifications.

\begin{table}[ht]
\centering
\begin{tabular}{@{}llrr@{}}
\toprule
\multicolumn{2}{l}{}
&Mean&Median\\ \midrule
\multirow{2}*{Recall}&Before&0.898&1\\
&After&0.902&1\\
\multirow{2}*{Jaccard distance}&Before&0.645&0.667\\
&After&0.627&0.667\\
\multirow{2}*{Edit distance}&Before&6.384&5\\
&After&6.397&5\\
\bottomrule
\end{tabular}
\Description[A table of scores]{A table of mean and median scores of the recall rate, Jaccard distance and edit distance before and after the modification}
\caption{Mean and median value of recall, Jaccard distance, and edit distance between user-provided focus and system-estimated focus model. See the caption of Figure \ref{fig:concepts_distribution} for details.}
\label{tab:mean_medium}
\end{table}

The mean of recall is around 0.9 and the median of recall is 1. This means that our Focus Model could capture almost all the concepts described by participants. 
We find three reasons for those uncaptured concepts. 
First, the participant just hovers over the concept, so even if they focus on this concept, our model can't capture it. We discuss this issue and its solution in \S\ref{sec:discussion} later in the paper.
Second, the participant gets inspiration from the Abstract List, but they can't find these concepts in the visualization interface especially the outcome scatterplot. This actually means the participant gets new findings from existing literature beyond the information provided by the data set. \S\ref{sec:interview} talks on the responses we received from these participants during the interview time.
Finally, the participant considers related attributes beyond the medical naming system used in the Focus Model. This usually implies that the participant has prior knowledge to do more reasoning in the analysis. 
For example, a participant mentions \textit{"physical pain"} in the self-reported concept list, and in our Focus Model we have diagnosis \textit{"M86.10: Other acute osteomyelitis, unspecified site"} as wells as \textit{"S09: Other and unspecified injuries of head"}. 
Both osteomyelitis (an infection of the bone) and head injury are accompanied by some kind of physical pain. However, as concepts are from different perspectives we do not map \textit{"physical pain"} to the two codes. 

We analyze the Jaccard distance and edit distance of the comparison to show the accuracy of our Focus Model. Although we do not obtain any baseline due to the novelty of representing analytic focus, our result could provide references to future studies. 

We then measure how our Focus Model's performance change with time. In all of the 24 participants, 7 have 4 pauses, 15 have 3 pauses and 2 have 2 pauses. We leave out the 2 participants who only have 2 pauses and split the other participants into two groups. We calculate the average scores of all three measurements for each round separately for each cohort. The data are shown in Table~\ref{tab:percentage_3round}.

\begin{table}[ht]
\centering
\begin{tabular}{llrrr|rrrr}
\toprule
\multicolumn{2}{l}{}
&Round1&Round2&Round3 & Round1&Round2&Round3&Round4\\ \midrule
\multirow{2}*{Recall}&Before&0.985&0.894&0.986 &0.833&0.834&0.869&0.670\\
&After&0.987&0.906&0.986 &0.833&0.849&0.869&0.670\\
\multirow{2}*{Jaccard distance}&Before&0.627&0.642&0.638 &0.554&0.543&0.625&0.830\\
&After&0.578&0.635&0.630 &0.523&0.527&0.625&0.830\\
\multirow{2}*{Edit distance}&Before&4.308&5.000&7.000 &4.000&7.000&8.714&12.714\\
&After&4.000&5.286&7.071 &4.140&7.000&8.714&12.714\\
\bottomrule
\end{tabular}
\Description[A table of scores]{A table of scores of the recall rate, Jaccard distance and edit distance in each round for participants with 3 and 4 pauses, before and after the modification}
\caption{The average recall, Jaccard distance, and edit distance  in each pause round. Columns on the left side of the dividing line are for participants with 3 pauses; columns on the right side of the dividing line are for participants with 4 pauses. We separate the scores before and after the participant's modification. Recall: the larger the better; Jaccard distance and edit distance: the smaller the better.}
\label{tab:percentage_3round}
\end{table}

In general, our Focus Model's performance is accurate and stable over time. We notice that the edit distance has an increasing trend in both the 3-round test and 4-round test. While the recall rate is stable, this indicates that the order accuracy decreases over time. A few participants mention this in the interview that they also notice the same issue (\S\ref{sec:interview}). There are multiple reasons to explain this, for example, parameters are not adapted to any individuals, and we discuss solutions in \S\ref{sec:discussion}.
In addition, there is an abrupt drop-down for recall between the third round and fourth round in the 4-round test. We found that 20 minutes usually let the participants wrap up a conclusion for the task, with the extra time, they might start a new sub-task from the start. Our Focus Model could not capture and divide sub-tasks in analysis, and we provide possible methods to solve this problem in \S\ref{sec:discussion}. Participants who have 4 round data-collecting opportunities get familiar with Cadence and the data set more quickly than the other group. They are better at utilizing the functionalities we provide. The three people we mention in \S\ref{sec:interview} who use terms inspired by the abstracts to search in the data set are all belong to this 4-round group. Hence we get the abrupt drop-down for recall in the fourth round.

\subsubsection{Post-task Questionnaire}

\begin{figure}[ht]
\centering  
\includegraphics[width=0.6\textwidth]{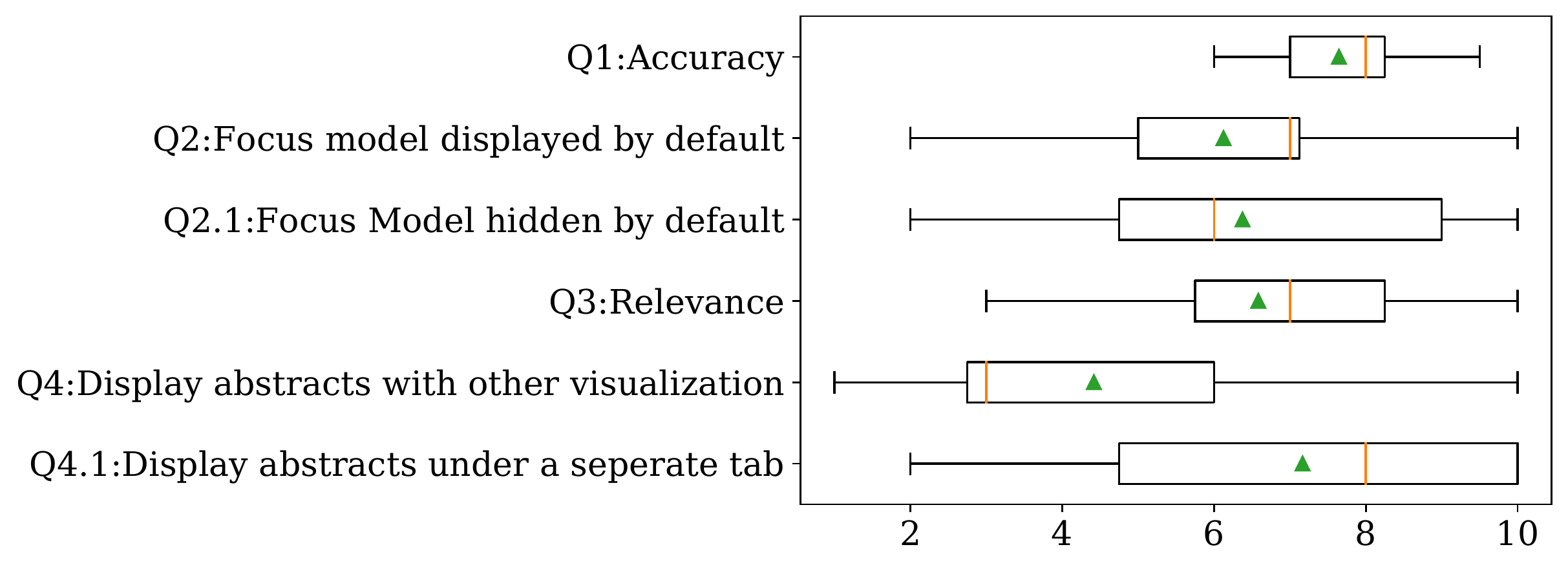}
\Description[A box plot with error bars]{A box plot showing the score distributions of the six questions, with average score markers}
\caption{Box plots for all six questions in the post-task questionnaire. Green triangles mark the average; vertical orange bars mark the median; two ends of each box mark the 25\% and 75\% percentiles; two ends of each whisker mark the maximum and minimum values. }
\label{fig:questions_distribution}
\end{figure}

In the post-task questionnaire, we ask 6 questions for ratings based on a scale from 1 to 10 with 1 as the most negative rating and 10 as the most positive rating. Q1 is on the accuracy of the Focus Model; Q2 and Q2.1 are on the display preferences of the Focus Model, visible along with all the other visualizations or hidden by default; Q3 is on the relevance of surfaced abstracts; Q4 and Q4.1 are on the display preferences of the Abstract List, visible along with all the other visualizations or hidden by default. Figure~\ref{fig:questions_distribution} shows the result including averages of scores for all six questions from all 24 participants.

Overall, participants think that our Focus Model is relatively accurate with an average score of 7.6 and a medium score of 8 for Q1. For the general helpfulness of the Focus Model, 21 participants give a score larger than or equal to 5, indicating that most participants find the Focus Model helpful to their analyses. The average scores of Q2 and Q2.1 are 6.13 and 6.38. For the individual preference of displaying or hiding the Focus Model by default to maximize its helpfulness, 9 people prefer displaying, 12 people prefer hiding, and 3 people think both are of equivalent helpfulness. 

For the Abstract List relevance, participants think it is relevant but with more improving space, with an average score of 6.6 and a medium score of 7 for Q3. As of the general helpfulness of the Abstract List, 20 participants give a score larger than or equal to 5, meaning that most participants find the Abstract List helpful to their analyses. The average scores of Q4 and Q4.1 are 4.4 and 7.2. This also indicates that more participants prefer the Abstract List under a separate tab by default. For the individual preference of displaying or hiding the Abstract List by default to maximize its helpfulness, 7 people prefer displaying, 16 people prefer hiding, and only 1 person thinks both are of equivalent helpfulness.

\subsubsection{Common answers from the interview}\label{sec:interview}

After the post-task questionnaire, we also conduct a short semi-structured interview for each participant. In the interview, we ask the reasons for the scores they give in the questionnaire, and based on their answers we may follow-up with personalized questions. We conclude the reasons that have commonalities to explain the scores in this section. We talk about other interesting or inspiring findings from our interviews in \S\ref{sec:discussion}.

First, we ask all participants \textbf{how accurate they think our Focus Model is and why}. Most people express that the accuracy in terms of medical concept capture is high enough to capture all or most of their analytic focus and the order of these concepts is similar to their subjective ranking order. 
Three mention that the Importance Scores were not sensitive enough to capture recent concepts but putting too many weights on earlier, stale concepts. We consider this as a combination of focus switch detection problem and parameter tuning problem (see more in \S\ref{sec:discussion}).

Second, we ask \textbf{how helpful they think the Focus Model is for the data analysis}. In addition, we also let them provide reasons for their display preferences based on the scores to Q2 and Q2.1. Overall, almost all participants believe that the current Focus Model could be a great log reference for their analysis, and as a possible supplement or replacement for tedious note-taking. As of the usage of this log reference, participants diverge into two groups: (1) those that feel it is most useful to review the focus live during their analysis as well as after completing their work, and (2) those who feel it is most useful to view the focus model after the analysis has been completed.  Interestingly, the participants that are professional analysts identified most with the second group, suggesting that their analysis should be driven by data and their own knowledge.  They suggest that a manual review of the focus model is useful after the analysis has been completed as a way to assess their work.  We note that the design of the model to decay the importance of concepts over time suggests that this approach would be most beneficial at the scale of relatively small units of work where the focus persists at the end of a task.

This also reflects the diverge of display preferences. The common reason for checking the Focus Model during the analysis is that it can remind them of concepts they once explored but forgot. About five people point out that this is valuable especially when they analyze an unfamiliar topic since the concepts explored can be forgotten more quickly. The main reason to check the Focus Model after the analysis is that they can retrospectively check the Focus Model after the analysis to see if anything is omitted in the conclusion, even after several months for some additional analysis. People also express their worries about putting the Focus Model besides other graphs and checking it during the analysis: the information in the Focus Model is circulatory to the things in mind so this may strengthen possible biases; the user interface may be too crowded and the Focus Model may distract the user. One participant also mentions that such logs can be leveraged in collaborative works, that others could check the log to pick up things missed in the analysis, or even continue the analysis if the system could read the Focus Model and recover the graphs where left.

Third, we ask \textbf{participants who have modified their Focus Description Questionnaires after checking the Focus Model result at least one time why they have modified the questionnaire}. In total 15 participants have modified the questionnaire. All of them explain that they forget the concepts and the Focus Model reminds them. This proves that our Focus Model could capture things beyond people's memory and help remind the user.

Last, we also ask \textbf{how helpful the Abstract List is to the analysis}. Overall, participants agree that the Abstract List is helpful to capture new ideas and strengthen their current analysis, although a few of them also state that some of the retrieved abstracts sometimes appear to lack relevance with their focus. Three people mention that these abstracts bring out new relative concepts that inform subsequent search back in the original data set. One searches for \textit{Myocardial Infraction}, hinted from the Frequently used terms, in the outcome scatterplot. This literature can also provide background information of the field to help understand the data, and affirm more complicated relationships like published evidence for causal relationships among concepts that align with simple correlations found in the data visualization. 
Although one analyst mentions that the Abstract List result could potentially be biased compared to a full literature review, another former data analyst who worked in the medical consulting field expresses a lot of appreciation for the approach. This analyst reports regularly using dual monitors to display Tableau and PubMed separately while doing analysis. The participant suggests that the approach in our study would save time and increase the efficiency of the work. A future study could directly compare the approach outlined in this paper with this analyst's typical siloed analysis environment.

%% file: tex/discussion.tex
\section{Discussion}\label{sec:discussion}

In this section, we will talk about the limitations and potential benefits of the Focus Model and Abstract List according to our observations and results of the user study. 

\subsection{Calculating the Importance Score}

The prototype of the Focus Model shows relatively high accuracy in the user study, however, there exist limitations for our current implementation and we propose methods to improve it. We do not capture the mouse hover as a semantic action. However, given the feedback from several participants, hover over certain seconds may indicate that the user is currently interested in this object. We could assign a lower Persistence Score and Importance Score to such action. 

Some participants find that in later rounds, concepts of their previous focus might still maintain a high Importance Score and the latest concepts are lower in the rank. We observe that this usually happens when the user starts to clear out components, indicating that the user is switching to another part of the task. We could adopt such a "sub-task" conception in our model to reflect the analysis process better. In data analysis, it is common to break a task into multiple sub-tasks and combine the conclusions to solve a problem. Each sub-task corresponds to a Focus Model. 
We could introduce a task-switching mechanism in the current Focus Model Engine, however, it will be challenging to determine where to split sub-tasks given a series of actions. A first-step idea is binding a sub-task Focus Model to a cohort panel.
Psychology studies~\cite{barrouillet2004time} on task-switching working memory spans may help develop the model to recognize sub-tasks in data analysis.

We also receive feedback on leveraging the existing hierarchies in computing the scores and presenting concepts in the Focus Model in Cadence. ICD-10 provides a systematic method to categorize any diagnosis in a hierarchy. If the concepts in the Focus Model have overlaps in definition, treating them as independent concepts and assign scores separately might extra emphasize these related concepts. We can also embed such a hierarchy in presenting the Focus Model to users.

During the user study, we observe that a few participants use the functions provided by Cadence differently from how they are designed. For example, one participant adds a series of nested milestones in a single timeline rather than using the filter.
The person will reset the timeline and add some of the milestones back to remove some milestones. 
As a result, those concepts added multiple times have a higher importance score than the newly added one which is usually the most important in the person's focus. This inspires us that we could develop a self-adapted parameter assignment system based on the user's habit. This also enables a more personalized Focus Model in the future.

\subsection{Concept Granularity}

The feedback received via the Focus Description Questionnaire provided several interesting subjective observations related to concept granularity which provide useful insights for improving our proposed modeling approach in the future.
More specifically, while the Focus Model itself represents concepts as a specific diagnosis or procedure (or category of diagnosis/procedure) based on the formal coding system used by the system (e.g., ICD-10), people often focus on things in a less rigid way.  

First, relationships between concepts, reflected in the questionnaires as words like "after" or "containing," were an important piece of how users verbalized their own analytic focus. In particular, given the focus in Cadence on time-sequenced analysis, about half of the participants use temporal relationship words to constrain their focus.
This suggests future extensions for temporal and relational modifiers within the Focus Model would be beneficial.

Second, we found that the same concept could mean different things to different people. For example, two participants reported the concept \textit{"physical pain"}. One user was referencing the specific ICD-10 concepts indicating pain, while the other used the same term to capture a broader range of diagnoses from the respiratory system to the circulatory system.  Similarly, the concepts reported in the questionnaire were not always used consistently. Several participants mentioned "injury" as a concept. However, after letting them explain if they meant the formal \textit{Injury (S00-T14)} category in ICD-10, some agreed while others did not. Those who disagreed usually explained that they meant a more specific concept of injury, such as \textit{shoulder injury} or \textit{joint injury}. Sometimes, this difference in granularity can be captured using the existing ICD-10 hierarchy, but other times (especially when a user has strong knowledge of the subject) the user may think about concepts according to other types of relationships beyond those formally present in the data representation.

\subsection{Assumption and Limitation of Focus Modeling Algorithm}
\label{sec:limitations}

The proposed approach to modeling analytic focus provides a general framework. Its core assumption is that a system can represent users' visual analytic activity in the form of discrete actions represented in the form of $(type, parameters)$ tuples as defined earlier in this paper. The set of $actions$ is system-specific -- determined by the views and controls that a system supports. The parameters for each action include $concepts$ which are application-specific -- determined by the data schema of a specific problem. Key parameters in the algorithm also need system-specific tuning, reflecting that the salience of individual actions each system implements is not universal.  Therefore, configuring the persistence scores, importance scores, and importance bias scores is important during implementation to produce the best possible results.

The algorithm can be instantiated in many analytics systems beyond Cadence. For instance, a geovisualization system can use it to model a user's focus when exploring environmental data. Actions may include \textit{zooming in/out}, \textit{panning}, \textit{adding/removing a layer}, and \textit{selecting individual items} on the map. In another example, a text search engine can use the proposed approach to model a user's focus during exploratory text search. Actions may include \textit{launching a query}, \textit{revising a query}, \textit{clicking on a result}, and \textit{clicking on a recommended query}.  In this direction, our lab is currently in the early stages of designing a search-based experiment that leverages the methods proposed in this paper.

Although the $(action, concept, time step)$  abstraction is general, it is not universally applicable. As previously described, it assumes that a discrete sequence of actions will be reported and that the model should be updated after each new report is received.  The discrete nature of these actions is not compatible with some continuous interaction models such as those that might be used in continuously evolving visualizations of live streaming data.  Conversely, visualization systems that don't support exploratory interactions (such as static dashboards that are passively consumed) are not compatible.  Finally, while data dimensions can often map directly to concepts in the proposed model, it may be difficult in some systems to directly map user focus to different concepts.  For example, low-dimensional data visualizations with very few dimensions may support interactions that largely center around changes in-range values for the same small number of dimensions rather than differences in concepts.  The model could be extended to support values in addition to concepts as action parameters, but it has not been the focus in the work presented in this paper. 

%% file: tex/conclusion.tex
\section{Conclusion}

In this paper, we studied the problem of modeling a user's analytic focus in an interactive exploratory data analysis. We designed computational approaches to representing and updating such a Focus Model, and implemented these approaches in an existing visual analytics platform for structured health data. As a direct application of the focus model, we leveraged it to bring relevant medical articles to the user's focus and complement the visualized findings in structured data.
We evaluated the prototype implementation in a user study. The results suggest that our proposed algorithm was able to capture most of the concepts in the participant's focus. The vast majority of participants also find the retrieved medical articles are helpful in their analysis tasks.
Through qualitative data analysis, we identified many concrete directions for our research in the next step. Overall, our study shows the feasibility of modeling users' focus during exploratory visual analysis and leveraging it to broaden the exploration into other relevant data sources.